\documentclass[runningheads]{llncs}
\usepackage[T1]{fontenc}
\usepackage{graphicx}
\usepackage{graphicx}
\usepackage{booktabs}
\usepackage{multirow} 
\usepackage{hyperref} 
\usepackage{subcaption}
\usepackage{caption} 
\usepackage{rotating,tabularx}
\usepackage{xcolor}

\begin{document}
\title{Assessing the generalization performance of SAM for ureteroscopy scene understanding}
\titlerunning{Assessing the generalization performance of SAM for ureteroscopy scene}
%


\author{Martin Villagrana\inst{1}$^{,\dagger}$ \and
Francisco Lopez-Tiro\inst{1,2}$^{,\dagger}$ \and
Clement Larose\inst{3} \and \\
Gilberto Ochoa-Ruiz\inst{1}  \and
Christian Daul\inst{2}}


\authorrunning{M. Villagrana et al.}
%

\institute{Tecnologico de Monterrey, School of Engineering and Sciences, Mexico 
\and
CRAN UMR 7039, Université de Lorraine and CNRS, Nancy, France \and
CHRU de Nancy-Brabois, service d'urologie, Vandœuvre-les-Nancy, France \\ $^{\dagger}$Equal contribution \\ $^{*}$Corresponding authors: \email{gilberto.ochoa@tec.mx, christian.daul@univ-lorraine.fr}
}

\maketitle              
\begin{abstract}
The segmentation of kidney stones is regarded as a critical preliminary step to enable the identification of urinary stone types through machine- or deep-learning-based approaches. In urology, manual segmentation is considered tedious and impractical due to the typically large scale of image databases and the continuous generation of new data. In this study, the potential of the Segment Anything Model (SAM)—a state-of-the-art deep learning framework—is investigated for the automation of kidney stone segmentation. The performance of SAM is evaluated in comparison to traditional models, including U-Net, Residual U-Net, and Attention U-Net, which, despite their efficiency, frequently exhibit limitations in generalizing to unseen datasets. The findings highlight SAM’s superior adaptability and efficiency. While SAM achieves comparable performance to U-Net on in-distribution data (Accuracy: 97.68 ± 3.04; Dice: 97.78 ± 2.47; IoU: 95.76 ± 4.18), it demonstrates significantly enhanced generalization capabilities on out-of-distribution data, surpassing all U-Net variants by margins of up to 23\%.

\end{abstract}

\section{Introduction}
\label{sec:intro}

Kidney stone formation is recognized as a prevalent condition affecting millions of individuals globally. In the United States, lifetime prevalence rates have been reported to range from 7.2\% to 10.1\%, while higher rates—such as 21.1\%—have been observed in specific populations \cite{moftakhar2022prevalence}. Early detection and accurate diagnosis of kidney stone formation are essential to ensure the prescription of effective treatments and to mitigate complications, including renal damage, infections, and recurrent stone formation \cite{chew2024prevalence}. Through timely intervention, patient outcomes can be significantly improved, and healthcare costs may be reduced. Due to its high incidence and association with severe pain, the detection of kidney stones is considered a critical component of medical practice, particularly for patients with recurrent episodes or those at elevated risk \cite{daudon2018recurrence}.

The current diagnosis of kidney stones is primarily based on Morpho-Constitu-tional Analysis (MCA), a process that depends extensively on visual inspection and biochemical analysis performed by clinicians \cite{daudon2004clinical}. Consequently, numerous attempts have been made to support this diagnostic process through computer vision-based approaches. Recent advances in deep learning (DL) have demonstrated promising potential for improving kidney stone classification \cite{lopez2024vivo}, \cite{estrade2022deep}, \cite{mendez2022generalization}. These methods could accelerate diagnosis by automating the traditionally operator-dependent MCA when conducted during endoscopic stone recognition (ESR) procedures \cite{estrade2022towards}, a critical advancement for patients requiring immediate intervention. Accurate segmentation of detected kidney stones is regarded as an essential preliminary step prior to classification, as it allows the stone to be isolated from surrounding tissues, thereby enhancing diagnostic accuracy \cite{ghosh2018effective}. Deep learning techniques, particularly convolutional neural networks (CNNs) and transformer-based models, have exhibited notable efficacy in segmentation tasks, as relevant features within complex medical images can be automatically identified, improving the reliability of image-based diagnosis \cite{lopez2021assessing}, \cite{gupta2022multi}.

Traditional segmentation models, including U-Net and its variants (Attention U-Net and Residual U-Net), have demonstrated strong performance in image segmentation tasks. The U-Net architecture, while recognized for its limited data efficiency, has been shown to effectively preserve spatial information through its encoder-decoder structure, though challenges remain when processing complex or noisy images \cite{ronneberger2015u}. Residual U-Net, which incorporates residual connections, has been designed to mitigate issues such as vanishing gradients, thereby enhancing performance in deeper network architectures; however, its increased depth can result in longer training times \cite{alom2019recurrent}. Attention U-Net enhances traditional U-Net through the integration of attention mechanisms, enabling the model to concentrate on diagnostically relevant regions, though this improvement comes at the cost of higher computational requirements \cite{li2022residual}. More recently, transformer-based models like the Segment Anything Model (SAM) \cite{kirillov2023segment} have been developed, offering notable adaptability across diverse image types without necessitating extensive task-specific modifications. Despite SAM's promising generalization capabilities, its implementation remains constrained by computational complexity and significant resource demands.

\begin{figure}[t] 
    \centering
    \subfloat[Traditional kidney stone segmenting method.]{\label{fig:traditional}
    \includegraphics[width=0.95\textwidth]{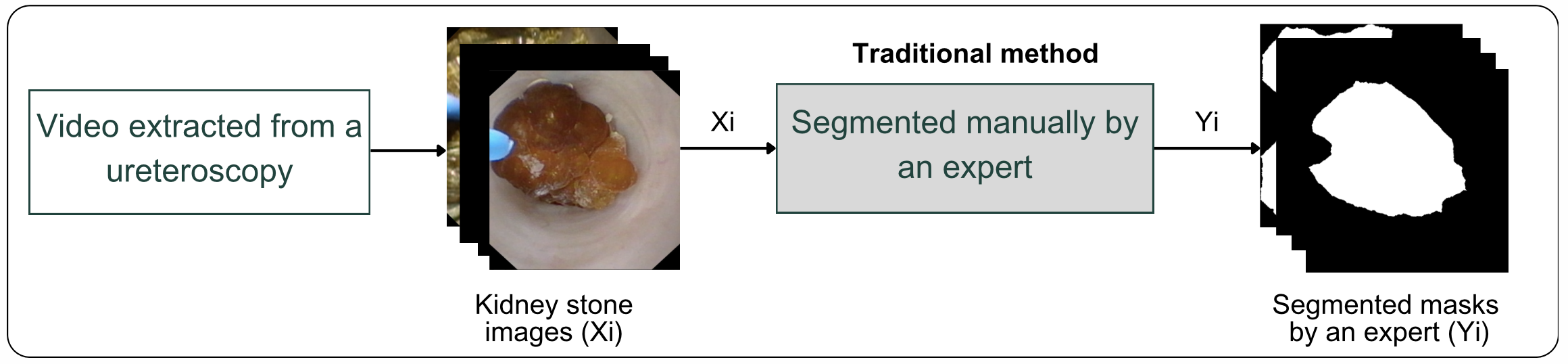}}
    \vspace{1mm}
    
    \subfloat[Automatic kidney stone segmentation and generalization method.]{
\label{fig:automatic}\includegraphics[width=0.95\textwidth]{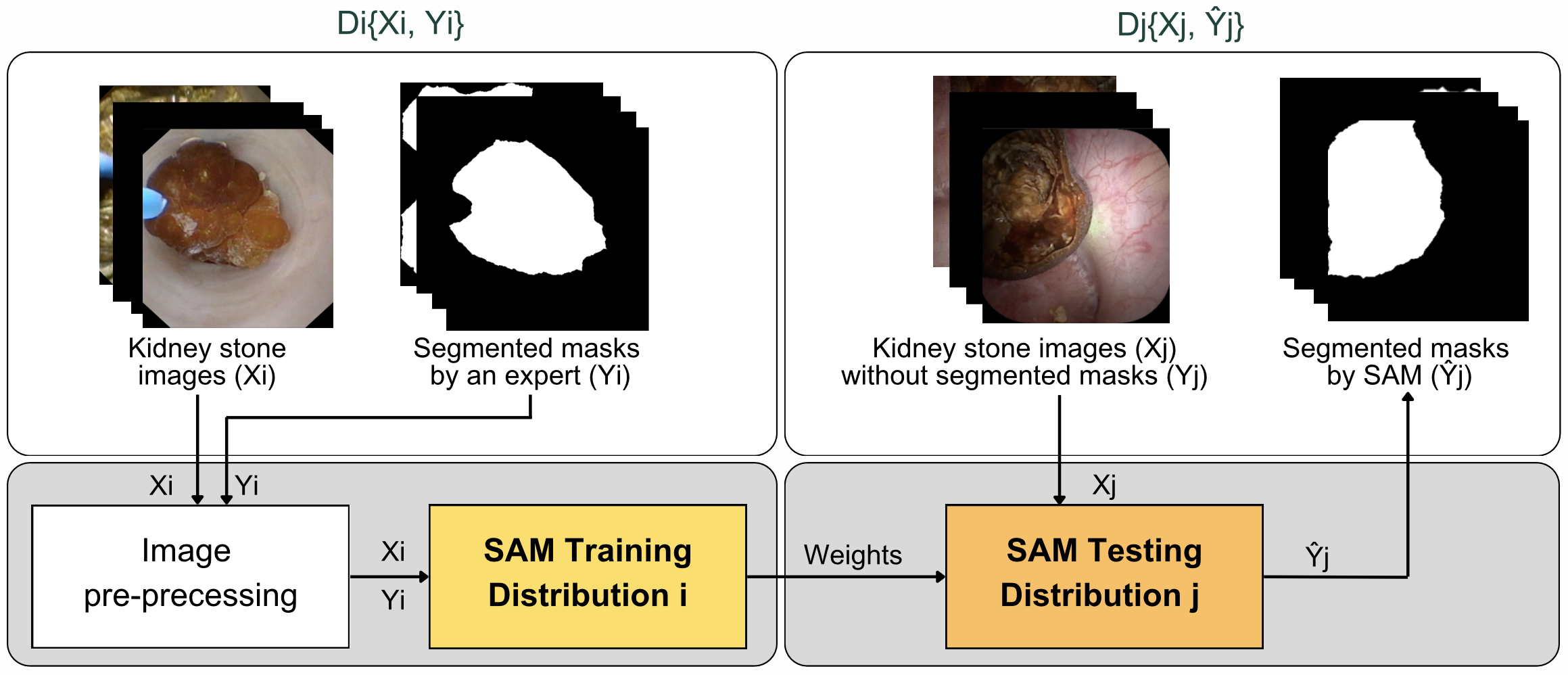}}

    \caption{Comparison framework for kidney stone segmentation methods. In traditional approaches (top), frames are extracted from ureteroscopy videos to create Dataset $D{_i}$ (distribution i), followed by manual expert labeling. The Segment Anything Model pipeline (bottom) is trained on \textit{i} and its labels, then performs inference on a distinct unlabeled Dataset \textit{j} (distribution j). Critical note: $D{_i}$ and $D{_i}$ represent different data distributions.} 
    \label{fig:method}
    \end{figure}

The segmentation of kidney stone images obtained from ureteroscopy videos has traditionally been conducted manually by clinical experts (Fig. \ref{fig:traditional}). While this approach has proven effective for limited datasets, its application to large-scale databases has been recognized as impractical due to the time-consuming nature of the process and its inherent dependence on operator skill \cite{estrade2022towards}. These limitations could potentially be addressed through advanced deep learning models like SAM, which have been shown to enable accurate and efficient automated segmentation of extensive video collections \cite{kirillov2023segment}. SAM's demonstrated capability to generalize across heterogeneous datasets represents a substantial technological advancement, positioning it as a viable alternative to conventional manual segmentation methods for large-scale applications (Figure \ref{fig:automatic}).

The principal contributions of this work are summarized as follows:

(i) Demonstrated Superior Generalization of SAM. 
The Segment Anything Model is shown to exhibit exceptional generalization capabilities for kidney stone segmentation tasks, systematically outperforming traditional architectures (U-Net, Residual U-Net, Attention U-Net). While conventional models exhibit significant performance degradation when applied to unseen data, SAM maintains robust accuracy across both known and unknown distributions when trained on identical datasets. This advancement highlights SAM's potential as a reliable tool for renal image analysis under real-world clinical variability.

(ii) Validated Multi-Class Adaptability Without Retraining. 
A novel three-class SAM variant ("kidney stone," "laser fiber," and "surrounding tissue") is validated, achieving high performance across both known and unknown distributions—a first in renal segmentation literature. Crucially, this adaptability is accomplished without requiring model retraining, as precision is preserved for both two-class and three-class tasks. This flexibility addresses a critical clinical challenge, where segmentation needs often vary across procedures and patient-specific anatomical presentations.

This work is organized as follows: Section \ref{sub:materials-methods} details the experimental dataset and methodological framework, with particular emphasis on comparative analysis between conventional deep learning-based segmentation approaches and the SAM. Section \ref{sec: Results} presents and analyzes the experimental findings, highlighting key performance comparisons. The study concludes with future research directions and final remarks in Section \ref{sec: Conclusions}.

\section{Materials and Methods}
\label{sub:materials-methods}

\subsection{Clinical image datasets}
\label{sec:data}
Four distinct kidney stone datasets were utilized in our experimental framework (Fig. \ref{fig:example_dataset}). The image data were acquired through two primary modalities: (1) conventional CCD camera systems and (2) endoscopic imaging captured during ureteroscopic procedures. The key characteristics of these datasets are detailed as follows:

\begin{figure}[ht]
    \centering
    \includegraphics[width=0.8\linewidth]{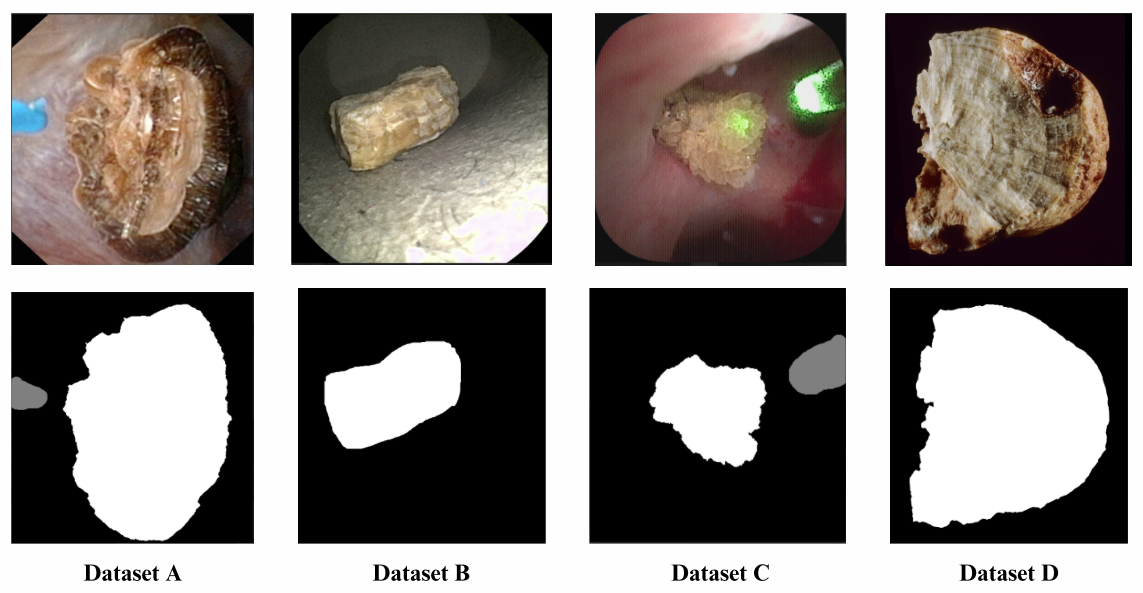}
    \caption{ The four datasets (distributions) of kidney stone images are displayed alongside their corresponding segmentation masks, which include the stone and laser fiber. From right to left: Dataset A (in vivo endoscopic), Dataset B (ex vivo endoscopic), Dataset C (in vivo endoscopic), and Dataset D (ex vivo CCD camera). All datasets enable two-class segmentation (kidney stone and tissue), while only Datasets A and C include a third class (laser).}
    \label{fig:example_dataset}
\end{figure}

\textbf{Database A  ($D_{A}$): In-vivo Endoscopic Images.} 
This dataset comprises 156 in-vivo endoscopic images capturing three distinct classes: (1) surrounding tissue, (2) kidney stone, and (3) laser fiber. While designed for either two-class (kidney stone and tissue) or three-class segmentation tasks, only 44 images include complete three-class annotations (with the additional "laser" class). All images feature variable dimensions, typically approximating 1008 × 1042 pixels \cite{estrade2022deep}.


\textbf{Database ($D_{B}$). Ex-vivo endoscopic images.}
This dataset contains 409 ex-vivo images acquired through a simulated renal environment using a kidney cavity phantom. The images feature two annotated classes: kidney stone and surrounding tissue. All images maintain a consistent resolution of 1920$\times$1080 pixels \cite{el2022evaluation}.

\textbf{Database C ($D_{C}$). In-vivo endoscopic images.}
This dataset comprises 138 clinical in-vivo images extracted from procedural videos, capturing three distinct classes: (1) living tissue, (2) kidney stones, and (3) surgical instruments (including laser fibers for stone fragmentation). The images are available in two resolutions:  400$\times$400  and 720$\times$720 pixels. This dataset can be used to evaluate the performance of segmentation mode for either two or three classes.
Dataset C remains under active clinical acquisition; consequently, all analyses were performed on the most contemporaneous available data stream.

\textbf{Database ($D_{D}$). Ex-vivo CCD-camera.}
This dataset consists of 356 ex-vivo images acquired using a charge-coupled device (CCD) imaging system, containing precisely two annotated classes: (1) kidney stone and (2) surrounding environment. The images exhibit variable dimensions, typically approximating 4288$\times$ 2848 pixels  \cite{corrales2021classification}.

\textbf{Data Partitioning and Preprocessing:}
The datasets were systematically partitioned with an 80\%--20\% training--testing split for Databases A, B, and D. Database C (153 images) employed a specialized division: 117 training images with two distinct test partitions (21 and 15 images), where test images were extracted from independent video sequences to ensure evaluation robustness.

All images were standardized to square dimensions of 
$512 \times 512$  pixels. Each database includes corresponding ground-truth segmentation masks, which were meticulously annotated by clinical specialists using the original high-resolution images for accurate performance evaluation.

\subsection{Kidney Stone Segmentation}
\label{sec: segmentation}

This work employs the Segment Anything Model for kidney stone segmentation in ureteroscopic images. The approach requires paired kidney stone images ($X_{i}$) and their corresponding ground-truth masks ($Y_{i}$) from a source distribution \textit{i}, formally denoted as  $D_{i}\left\{ X_{i},Y_{i} \right\}$. The segmentation masks are acquired through conventional manual annotation protocols, where kidney stone boundaries are meticulously delineated by clinical specialists (Fig. \ref{fig:traditional}).

Automated segmentation is subsequently performed using the SAM, through which new segmentation masks are generated across both the original training distribution ($D_{i}$) and previously unseen data distributions ($D_{j}$) (Fig. \ref{fig:automatic}).

\subsection{Traditional method for  segmenting kidney stones}

The segmentation masks ($Y_{i}$) are generated through a standardized protocol where multiple frames exhibiting kidney stones and adjacent tissue are first extracted from ureteroscopic video recordings. These frames are then meticulously annotated at the pixel level by urologists to delineate kidney stone, tissue, and instruments like basket or fiber laser.

 Finally, the annotated masks are systematically paired with their corresponding original images ($X_{i}$)  to form the complete dataset distribution $D_{i}\left\{ X_{i},Y_{i} \right\}$ for subsequent analysis.

\subsection{Automatic kidney stone segmentation and generalization} 
\label{sec:segmentation}

As previously established in Section \ref{sec:intro}, kidney stone segmentation has been demonstrated to be effective for limited datasets but becomes impractical when applied to large-scale collections. To address this limitation, the Segment Anything Model is evaluated using expert-derived annotations from the conventional manual segmentation protocol (Fig. \ref{fig:traditional}). Following training, the model is employed to automatically produce segmentation masks ($\hat{Y}_{i}$), with performance being assessed under two operational conditions: (1) application to the original training distribution ($D_{i} \rightarrow D_{i}$) and (2) generalization to previously unseen data distributions ($D_{i} \rightarrow D_{j}$) (Fig. \ref{fig:automatic}).

The Segment Anything Model implementation comprises two distinct phases: SAM-training and SAM-testing. During the training phase, an initial preprocessing stage is conducted where input images are normalized and resized to standardized dimensions of  $512\times512$ pixels, ensuring uniform data presentation across all model architectures.

In the SAM-training process, each input image is processed through a transformer-based feature extraction block, where it is condensed into a compact feature matrix. These extracted features are subsequently passed to a decoder head along with relevant model prompts, including potential segmentation masks. Following training completion, the optimized embeddings are utilized during SAM-testing to perform segmentation on either the original data distribution ($D_{i} \rightarrow D_{i}$) or alternative distributions ($D_{i} \rightarrow D_{j}$).

\subsubsection{Experimental setup.}  

This study systematically evaluates the Segment Anything Model's performance in automated mask generation across two distinct class configurations: (1) a two-class paradigm (kidney stone and surrounding tissue) and (2) a three-class paradigm (kidney stone, laser fiber, and tissue). For each configuration, comprehensive evaluations are conducted through both in-distribution and out-distribution experimental designs. The experimental framework is structured to assess model performance under the following conditions:

\begin{enumerate}
    \item[1.] Two-classes configuration. The first configuration focused on binary segmentation (kidney stone vs. tissue) applied to different kidney stone image domains (in-vivo and ex-vivo endoscopic, and ex-vivo CCD-camera). 

\vspace{1mm}
    
    \begin{itemize}
        \item \textit{In-distribution:} A single model was trained using SAM-training on dataset A (in-vivo endoscopic images) as the training domain. Once the model was trained with SAM, SAM-testing was performed on the test partition from the same distribution (in-distribution).

\vspace{1mm}
        
        \item \textit{Out-of-Distribution:} Subsequently, to evaluate the performance of the SAM-trained model on distribution A, inference was performed on the test partition from B, C, and D datasets. Notably, the model trained on A was neither modified nor adapted for these tests, making the evaluations on B, C, and D pure out-distribution cases.
    \end{itemize}
\end{enumerate}

Three U-Net variants—the standard model, Residual U-Net, and Attention-U-Net—were trained to benchmark SAM's performance against classical architectures. All models were trained exclusively on Dataset A and were subsequently evaluated on the test partition of A, B, C, and D datasets. This experimental design was implemented to specifically assess each architecture's out-of-distribution generalization capabilities.

The quantitative evaluation is based on standard metrics: Accuracy, Dice Coefficient, and Intersection over Union (IoU). These metrics can be used for an objective  segmentation quality assessment. In the metrics "Accuracy", "Dice", and "IoU", an upward arrow symbol $(\uparrow)$ is used to denote that higher values correspond to better performance. Conversely, in the "Error Rate" metric, a downward arrow symbol $(\downarrow)$ is employed to indicate that optimal results are those closest to zero.

Testing across four datasets (with distinct distributions) demonstrated the models' generalization capability and robustness for kidney stone segmentation across diverse clinical scenarios. A comparative analysis between SAM and U-Net-based models, including both in-distribution and out-of-distribution evaluations for the selected metrics, is presented in Table \ref{tab:two-classes}.

\begin{enumerate}
    \item[2.]  Three-classes configuration. The second configuration addresses three-class segmentation (kidney stone, laser, and tissue), utilizing exclusively Datasets A and C as these were the only datasets containing the 'laser' class annotation. %
    
Due to the limited amount of segmented data available in Dataset A for three classes (44 images), Dataset C (153 images) was selected for SAM initialization. Consequently, Dataset A was reserved exclusively for testing, while the model trained on C was employed for both training and evaluation purposes.
Although both datasets A and C were acquired in vivo using an endoscope, they are considered distinct domains. Each dataset was collected at different times with different instruments, resulting in varying image resolutions and quality.

A dual-model SAM training strategy was implemented: the first model was designed to segment kidney stone and tissue (replicating Experiment 1's approach), while the second model was specialized for laser and tissue segmentation. The predictions from both models were subsequently integrated through fusion to generate final three-class masks, enabling evaluation of SAM's performance on this more complex multiclass task.

\vspace{1mm}

    \begin{itemize}
        \item \textit{In-distribution:} For this training phase, the dataset C (comprising 117 training images) was used to initialize SAM-training. 
        The SAM testing procedure was conducted using two separate test partitions. The first test partition consisted of 21 images, while the second included 15 images, both derived from two different ureteroscopy videos. 

        \vspace{1mm}

        \item  \textit{Out-of-Distribution:} The performance of SAM, trained on distribution C and tested with the test partition of dataset A (44 images), was evaluated. It should be noted that the model trained on C was neither modified nor adapted for the test dataset.
    \end{itemize}
\end{enumerate}

The training of the models was performed on a 16 GB Nvidia DGX GPU,  using the AdamW Optimizer with a dynamic learning rate adjustment and a warmup phase. A combination of Dice Loss and CE Loss was used to minimize discrepancies between predicted segmentation and ground truth masks.  U-Net, Residual U-Net, and Attention U-Net were trained for 80 epochs, while SAM was trained for 200 epochs.

\section{Results and Discussion}
\label{sec: Results}

\begin{figure}[t]
    \centering
    \includegraphics[width=1\linewidth]{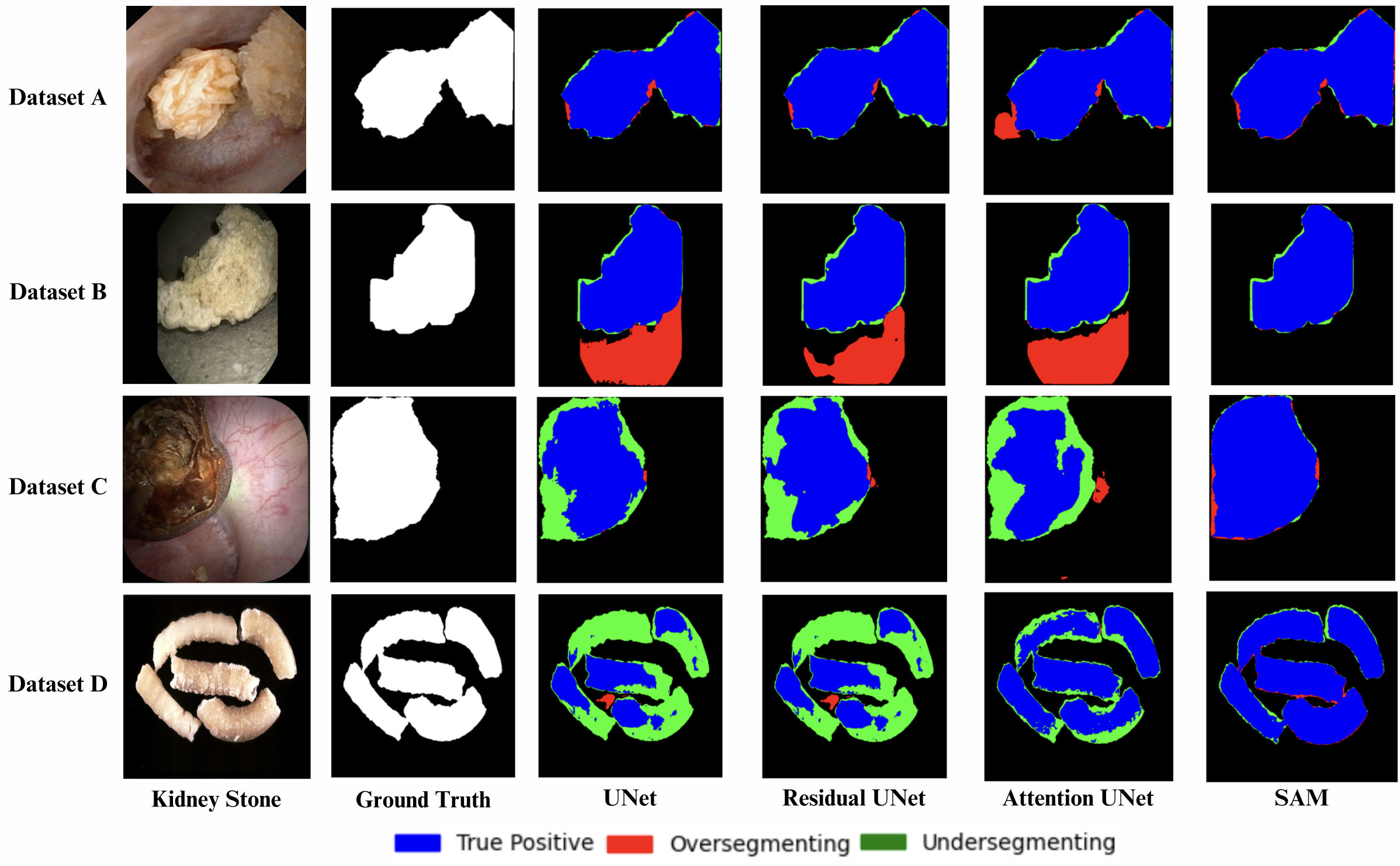}
    \caption{A qualitative comparison is presented across rows (Datasets A-D) and columns (kidney stone image, ground truth mask, U-Net, Residual U-Net, Attention U-Net, and SAM), where results are color-coded as: blue (true positives/correct segmentation), red (oversegmentation/false positives), and green (undersegmentation/false negatives).}
    \label{fig:comparative}
\end{figure}

The segmentation models (SAM and U-Net-based), described in Section \ref{sec:segmentation}, were evaluated using the datasets from Section \ref{sec:data}. Results were obtained for both two-class and three-class segmentation, including quantitative comparisons (Tables \ref{tab:two-classes} and \ref{tab:three-classes}) and qualitative evaluations (Figures \ref{fig:comparative} and \ref{fig:comparative_laser}). Quantitative analysis was performed using the Intersection over Union (IoU) metric, whose numerical values are reported in this section as performance indicators.

\begin{table}[t]
\centering
\caption{Comparison of the performance of four segmentation models (U-Net, Residual U-Net, Attention U-Net, and SAM) across four datasets (A to D). Segmentation is performed for two classes (surrounding tissue or background, and kidney stone). The presented results were measured using segmentation metrics (Accuracy, Dice Score, Intersection over Union, and Error Rate). The best results for each metric are highlighted in bold.
}
\vspace{2mm}
\label{tab:two-classes}
\begin{tabular}{@{}crcccc@{}}
\toprule 
Dataset                                & \multicolumn{1}{c}{Model} & Accuracy $\uparrow$    & Dice $\uparrow$         & IoU $\uparrow$          & Error rate $\downarrow$  \\ \midrule  \vspace{-0.0mm} 
\multirow{4}{*}{$D_{A}\rightarrow D_{A} $} & SAM Model                 & 97.68$\pm$03.04 & 97.78$\pm$02.47 & 95.76$\pm$04.18 & 4.24\%     \\ \vspace{-0.0mm}
                                       & U-Net                      & 97.68$\pm$01.38 & 97.72$\pm$01.28 & 95.77$\pm$02.37 & 4.23\%     \\ \vspace{-0.0mm}
                                       & \textbf{ResU-Net}                   & \textbf{97.92$\pm$00.96} & \textbf{97.93$\pm$01.01} & \textbf{95.97$\pm$01.91} & \textbf{4.03\%}     \\ \vspace{-0.0mm}
                                       & AttnU-Net             & 96.37$\pm$02.71 & 96.24$\pm$03.77 & 92.98$\pm$06.02 & 7.02\%     \\ \midrule \vspace{-0.0mm} 
\multirow{4}{*}{$D_{A} \rightarrow D_{B}$} & \textbf{SAM Model}                 & \textbf{98.71$\pm$02.87} & \textbf{96.39$\pm$07.70} & \textbf{93.74$\pm$09.55} & \textbf{6.26\%}     \\ \vspace{-0.0mm}
                                       & U-Net                      & 88.8$\pm$07.83  & 73.05$\pm$18.58 & 60.75$\pm$22.20 & 39.25\%    \\ \vspace{-0.0mm}
                                       & ResU-Net                   & 87.9$\pm$07.03  & 70.85$\pm$18.49 & 57.95$\pm$21.82 & 42.05\%    \\ \vspace{-0.0mm}
                                       & AttnU-Net             & 83.83$\pm$07.75 & 65.47$\pm$19.37 & 51.77$\pm$21.76 & 48.23\%    \\ \midrule \vspace{-0.0mm} 
\multirow{4}{*}{$D_{A}\rightarrow D_{C}$} & \textbf{SAM Model}                 & \textbf{95.33$\pm$04.70} & \textbf{92.42$\pm$07.19} & \textbf{86.64$\pm$10.90} & \textbf{13.36\%}    \\ \vspace{-0.0mm}
                                       & U-Net                      & 86.45$\pm$11.38 & 74.31$\pm$22.89 & 63.37$\pm$23.70 & 36.63\%    \\ \vspace{-0.0mm}
                                       & ResU-Net                   & 86.09$\pm$11.64 & 71.69$\pm$25.58 & 60.83$\pm$25.27 & 39.17\%    \\ \vspace{-0.0mm}
                                       & AttnU-Net             & 85.16$\pm$09.95 & 71.67$\pm$20.30 & 59.08$\pm$20.69 & 40.92\%    \\ \midrule \vspace{-0.0mm} 
\multirow{4}{*}{$D_{A} \rightarrow D_{B}$} & \textbf{SAM Model}                 & \textbf{96.48$\pm$07.91} & \textbf{96.50$\pm$06.20} & \textbf{93.81$\pm$09.47} & \textbf{6.19\%}     \\ \vspace{-0.0mm}
                                       & U-Net                      & 87.59$\pm$12.96 & 78.73$\pm$25.35 & 70.58$\pm$27.59 & 29.42\%    \\ \vspace{-0.0mm}
                                       & ResU-Net                   & 89.51$\pm$11.02 & 84.13$\pm$18.92 & 76.19$\pm$22.41 & 23.81\%    \\ \vspace{-0.0mm}
                                       & AttnU-Net             & 89.31$\pm$11.78 & 82.85$\pm$21.42 & 75.11$\pm$24.42 & 24.89\%    \\ \bottomrule
\end{tabular}
\end{table}

\subsection{Segmentation of kidney stone and surrounding tissue}
\label{sec:two-class}

\subsubsection{In distribution.}

The SAM model was trained on Dataset A for kidney stone and surrounding tissue segmentation, achieving 95.76 IoU (±4.18) on test data with consistent performance: 97.68\% accuracy (±3.04), 97.78\% Dice (±2.47), and 4.24\% error rate.

Similar results were observed for Distribution A ($D_A \rightarrow D_A$) in Table \ref{tab:two-classes} . High performance was demonstrated by SAM (95.76 IoU ±4.18), with comparable results being achieved by both U-Net and Attention U-Net. Although peak performance was not attained by SAM in same-distribution testing (Dataset A), competitive performance was maintained relative to Residual U-Net (97.97 IoU ±1.01).

Fig. \ref{fig:comparative} (first row) presents a qualitative comparison between SAM and U-Net-based segmentation methods. The visualization demonstrates that all models trained on Dataset A achieve near-perfect predictions within their training distribution: both oversegmented (red) and undersegmented (green) pixels are minimal compared to the true positive segmentation mask (blue).

Across all models, the error rates for in-distribution evaluation (Dataset A) remain relatively low: approximately 4\% for SAM, U-Net, and Residual U-Net, while Attention U-Net shows a slightly higher error rate of 7\%. However, out-of-distribution experiments demonstrate a significant increase in error rates due to the distribution shift.

\textbf{Out-of-distribution.} The SAM model, trained on Dataset A and evaluated across Datasets B, C and D, was demonstrated to achieve consistently superior performance compared to U-Net-based architectures in all evaluations. 

Specifically for Dataset B (ex vivo endoscopic images), an IoU of 93.74 (±9.55) was recorded, significantly exceeding the results obtained by: the standard U-Net 60.75 (±22.20), Residual U-Net 57.95 (±21.82), and Attention U-Net 51.77 (±21.76).
While standard U-Net maintained the second-highest performance at 60.75 (±22.20), it remained 32\% inferior to SAM in terms of the IoU metrics.
Fig. \ref{fig:comparative} (second row) qualitatively compares segmentation performance on out-of-distribution data: while U-Net-based models (trained on Dataset A) missegment kidney stones as surrounding tissue in Dataset B due to challenging illumination and phantom conditions, SAM maintains accurate segmentation, correctly distinguishing both classes.

In the evaluation of Model A on Dataset C (in vivo endoscopic images, different distribution), SAM achieved an IoU of 86.64 (±10.90), outperforming all U-Net-based models. The highest performance among U-Net variants was observed with standard U-Net at 63.37 IoU (±23.70), while Residual U-Net and Attention U-Net were found to have comparable results of 60.83 (±25.27) and 59.08 (±20.69) respectively. 
Regarding the qualitative results in Fig.  \ref{fig:comparative} (third row), the U-Net-based methods now correctly identify the target region. However, challenges like non-uniform illumination (similar to Dataset A) cause kidney stone undersegmentation (green pixels), particularly for edge pixels near the scene boundaries.

The final evaluation was performed on Model A trained using Dataset D (ex vivo CCD-camera images), which is characterized by fragmented kidney stones (post-extraction) in contrast to the intact stones typically found in Datasets A-C. Superior performance was demonstrated by SAM with an IoU of 93.81 (±9.47), followed by Residual U-Net [76.19 (±22.41)] and Attention U-Net [75.11 (±24.42)], whose results were found to be closely matched. The lowest performance was observed with standard U-Net at 70.58 (±27.59) IoU, representing a 23.23\% decrease relative to SAM.

As shown in Table \ref{tab:two-classes}, SAM trained on Distribution A and tested on in-distribution data maintains an error rate of 4.03\%, outperforming all U-Net-based models. Furthermore, SAM demonstrates robust out-of-distribution generalization across Datasets B, C, and D, consistently achieving both the highest performance in term of IoU and lowest error rates in all test scenarios.

\begin{table}[t]
\centering
\caption{Quantitative results of the SAM model trained on Dataset C (in vivo endoscopic dataset) for three-class segmentation (kidney stone, laser fiber, and surrounding tissue) are reported. Evaluation was performed on two distinct test partitions from Dataset C (corresponding to different video sequences) and additionally on Dataset A.}
\vspace{2mm}
\label{tab:three-classes}
\begin{tabular}{@{}lcccccc@{}}
\toprule
\multicolumn{1}{c}{Dataset} & Class         & Accuracy $\uparrow$    & Dice $\uparrow$        & IoU $\uparrow$         & Error Rate $\downarrow$ & Support \\ \midrule
                            &  \textbf{Laser} & 91.00$\pm$06.56 & 94.22$\pm$05.63 & \textbf{89.56$\pm$09.23} & \textbf{10.44\%  }  &         \\
\multicolumn{1}{c}{$D_{C} \rightarrow D_{C}$}   &  Stone & 91.00$\pm$06.56 & 87.84$\pm$06.65 & 78.93$\pm$10.30 & 21.07\%    & 21      \\
                            &  Mean           & 91.00$\pm$06.56 & 91.03$\pm$06.14 & 84.24$\pm$09.77 & 15.76\%    &         \\ \midrule
                            &  Laser & 84.03$\pm$10.81 & 83.60$\pm$11.81 & 73.58$\pm$17.37 & 26.42\%    &         \\
\multicolumn{1}{c}{$D_{C} \rightarrow D_{C}$}   &  \textbf{Stone} & 99.21$\pm$00.58 & 87.24$\pm$07.85 & \textbf{78.13$\pm$11.16} & \textbf{21.87\%}    & 15      \\
                            &  Mean           & 91.62$\pm$05.70 & 85.42$\pm$09.83 & 75.85$\pm$14.27 & 24.15\%    &         \\ \midrule
                            &  Laser & 99.70$\pm$00.21 & 87.87$\pm$16.98 & 81.11$\pm$18.49 & 18.89\%    &         \\
\multicolumn{1}{c}{$D_{C} \rightarrow D_{A}$}   &  \textbf{Stone} & 94.55$\pm$04.44 & 95.21$\pm$04.18 & \textbf{91.13$\pm$06.79 }& \textbf{8.87\% }    & 44      \\
                            &  Mean           & 97.12$\pm$02.33 & 91.54$\pm$10.58 & 86.12$\pm$12.64 & 13.88\%    &         \\ \bottomrule
\end{tabular}
\end{table}

\begin{figure}[ht]
    \centering
    \includegraphics[width=0.7\linewidth]{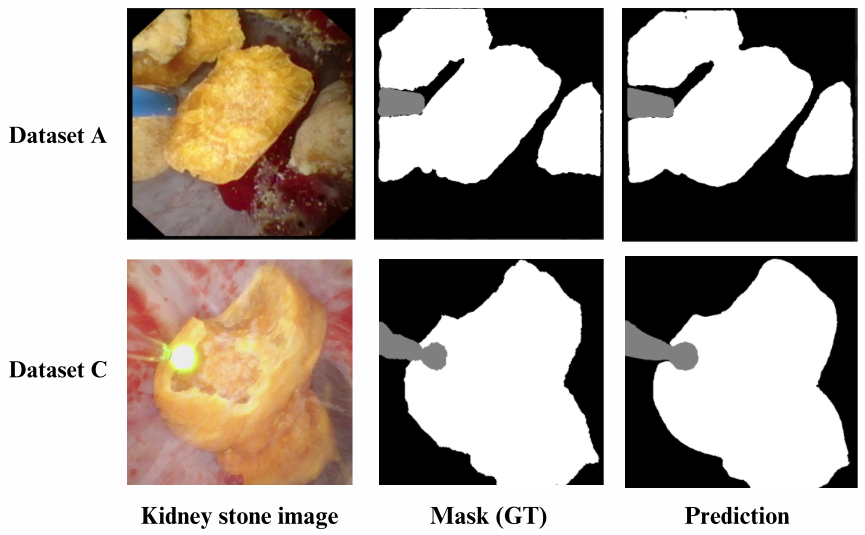}
    \caption{Qualitative comparison of segmentation results for three classes (kidney stone, laser, and surrounding tissue). From left to right: Kidney stone image, segmentation mask (ground truth), and the prediction generated by the SAM model trained on Dataset C. The first row corresponds to in-distribution results, while the second represents out-of-distribution performance.}
    \label{fig:comparative_laser}
\end{figure}

\subsection{Segmentation of kidney stone, laser fiber and surrounding tissue}
\label{sec:three-class}

Section \ref{sec:two-class} evaluated the performance of various segmentation models (SAM and U-Net-based architectures) for binary classification of kidney stones and surrounding tissue under both in-distribution and out-of-distribution conditions. A comprehensive analysis was conducted across all datasets described in Section \ref{sec:three-class}, as these datasets were fully annotated for binary segmentation tasks.

Section \ref{sec:three-class} now focuses exclusively on SAM's performance for in-distribution (Dataset C) and out-of-distribution (Dataset A) scenarios involving three-class segmentation (kidney stone, surrounding tissue, and laser). All results are reported using Intersection over Union (IoU) metrics to maintain consistency with prior analyses.

The quantitative results of this section are presented in Table \ref{tab:three-classes}, showing SAM's performance when trained on 117 images from Dataset C and evaluated across three test sets: (1) the first in-distribution partition (21 support images), (2) a second in-distribution partition (15 support images from a different video sequence), and (3) an out-of-distribution evaluation using 44 laser-containing images from Dataset A's 138-image collection.
The results are presented for the "kidney stone" and "laser" classes, with the mean value being calculated as the average of these two classes.

\subsubsection{In-distribution}

The following performance metrics were observed for the SAM model trained on Dataset C and evaluated on the first test partition (21 support images): an IoU of 89.56 (±9.23) with a 10.44\% error rate was achieved for the laser class, while a slightly lower IoU of 78.93 (±10.30) and 21.07\% error rate were recorded for the kidney stone class.

Superior performance was demonstrated by the model for the "kidney stone" class compared to the "laser" class when evaluated on the second test partition C (15 support images). An IoU of 78.13 (±11.16) was achieved for kidney stone segmentation, while a slightly lower IoU of 73.58 (±17.37) was recorded for laser detection. Similar error rates of approximately 20\% were observed for both classes.

\subsubsection{Out-of-distribution}

To determine the performance of the SAM model trained on Dataset C, the same model used in the in-distribution experiments was evaluated on a different distribution, Dataset A (support: 44 images).
It showed superior performance for the "Kidney stone" class, with a 91.13±6.79 IoU and an 8.87\% error rate, compared to the "laser" class, which achieved an 81.11±18.49 IoU and an 18.89\% error rate. Despite being tested on Dataset A and trained on Dataset C, it achieved even better performance than the in-distribution predictions on Dataset C.

Fig. \ref{fig:comparative_laser} qualitatively compares the segmentation predictions of the SAM model (trained on Dataset C) when evaluated on both in-distribution and out-of-distribution data. Strong performance is observed for both the "kidney stone" and "laser" classes across distributions.

Finally, the SAM model (trained on the in-vivo endoscopic dataset C) for three classes (kidney stone, laser fiber, and surrounding tissue) was used to evaluate the full sets A, B, C, and D for two classes (kidney stone and tissue), aimed at determining its performance when segmenting two-class images. Encouraging results were obtained: for dataset A (in-vivo endoscopic, out-of-distribution), an IoU of 91.60±8.49 and an error rate of 8.40\% were achieved. For dataset B (ex-vivo endoscopic, out-of-distribution), an IoU of 90.60±14.62 and an error rate of 9.40\% were recorded. The best performance in both metrics was observed in dataset C (in-distribution), with an IoU of 93.71±1.15 and an error rate of 6.29\%. Finally, for dataset D (Ex-vivo CCD-Camera, out-of-distribution), an IoU of 91.97±13.07 and an error rate of 8.03\% were attained.

\section{Conclusions}
\label{sec: Conclusions}

This study demonstrates that the two-class SAM model outperforms traditional segmentation models (e.g., U-Net, Residual U-Net, Attention U-Net) in kidney stone segmentation. While traditional models performed well on their training dataset, their accuracy significantly declined on unseen datasets. In contrast, SAM, trained on the same data, achieved robust performance on the original dataset and generalized effectively across diverse datasets, maintaining high segmentation precision.

For the three-class SAM model ("kidney stone, laser fiber, and surrounding tissue"), strong performance was observed both in-distribution and out-of-distribution. Notably, the model retained its accuracy when classifying both two and three classes without retraining.

SAM’s ability to generate precise, artifact-free segmentations—combined with superior cross-dataset generalization—establishes it as the most reliable model for kidney stone segmentation. These findings underscore the importance of adaptable models for clinical applications with inherent data variability.

\section*{Acknowledgements}
The authors acknowledge the support of the “Secretaría de Ciencia, Humanidades, Tecnología e Innovación” (SECIHTI), the French Embassy in Mexico, and Campus France through postgraduate scholarships, as well as the Data Science Hub at Tecnológico de Monterrey. This work was also funded by Azure Sponsorship credits from Microsoft’s AI for Good Research Lab under the AI for Health program and the French-Mexican ANUIES-CONAHCYT Ecos Nord grant (MX 322537). This work is also partially supported by the Worldwide Universities Network  (WUN) project Optimize: "Novel robust computer vision methods and synthetic datasets for minimally invasive surgery".


\bibliographystyle{unsrt}
\bibliography{references.bib}

\end{document}